\documentclass[reprint,aps,pra,floatfix,showpacs]{revtex4-1}

\usepackage{graphicx}
\usepackage{hyperref}
\usepackage{amsmath}
\usepackage{amssymb}
\usepackage{braket}
\usepackage{epstopdf}

\DeclareMathOperator{\Tr}{Tr}
\DeclareMathOperator*{\Motimes}{\text{\raisebox{0.25ex}{\scalebox{0.8}{$\bigotimes$}}}}
\newcommand\numberthis{\addtocounter{equation}{1}\tag{\theequation}}

\begin{document}

\title{Disappearance of macroscopic superpositions in perfectly isolated systems}

\author{Chae-Yeun Park}
\author{Hyunseok Jeong}

\affiliation{Center for Macroscopic Quantum Control, Department of Physics and Astronomy,\\
	Seoul National University, Seoul, 151-742, Korea}
\date{\today}

\begin{abstract}
Schr\"{o}dinger's illustration of an imaginary cat in a box, neither alive nor dead, leads to a question of whether and how long a macroscopic quantum superposition can exist in various situations.
It is well known that a macroscopic superposition is destroyed very quickly by environmental effects called decoherence.
On the contrary, it is often believed that a macroscopic superposition continues to ``survive'' if it is ideally isolated from its environment.
In this paper, using a well-established measure of macroscopic superpositions and  the eigenstate thermalization hypothesis,
we show that macroscopic superpositions even in ideally closed systems are destroyed by thermalization processes. 
We further  investigate specific examples of a disordered Heisenberg spin chain varied between the thermalization phase and the many-body localized (MBL) phase. This leads to consistent results; initial macroscopic superpositions disappear under thermalization while they may be preserved in the MBL phase in which thermalization does not occur.
\end{abstract}

\maketitle

The principle of quantum superposition is of crucial importance in quantum descriptions of physical systems.
It is a long-standing question whether quantum superpositions of a macroscopic scale can ever exist and why it is so hard to observe them \cite{schrodinger35}.
The decoherence program, based on openness of quantum systems, well explains that 
macroscopic superpositions, although they may have existed, must disappear very quickly due to their interactions with environments
\cite{zurek03}. 

It is often believed that if a macroscopic quantum superposition can be perfectly isolated as a closed system, it would not lose its properties  as a macroscopic superposition, i.e., Schr\"{o}dinger's cat in an ideally isolated box would not be ``killed.''
Of course, a quantum superposition in a closed system does not become a classical mixture as it does when environmental effects exist. 
However, it was  shown that 
expectation values of physical observables in isolated quantum systems
tend to evolve to averages over a thermal ensemble
due to the  thermalization process \cite{gogolin15,deutsch91,srednicki94,rigol08}. 
It is thus an intriguing and unexplored question whether and how long a perfectly closed macroscopic superposition can be preserved.
The answer to this question may deepen our understanding of quantum mechanics in a macroscopic limit.

Here, we address this question using a well-established measure for quantum macroscopicity
for multipartite spin systems~\cite{shimizu02,shimizu05,frowis12,kang16}.
The measure is equivalent to the maximum variance of  a macroscopic observable which is the sum of many local spin operators.
We first show that if an $N$-particle system  thermalizes, quantum macroscopicity after thermalization is bounded by the order of $N$. This means that a system cannot be in a macroscopic quantum superposition after thermalization. 
We find that the suppression of fluctuation of macroscopic observables relative to the system size, which is due to thermalization, leads to this result.
We also investigate many-body localized (MBL) systems which do not thermailze~\cite{anderson58,basko06,nandkishore14} and were recently realized in experiments~\cite{schreiber15,choi16,smith16}.
On the contrary to the thermalizing systems, our results using an effective model~\cite{serbyn13,huse14} shows that macroscopic superpositions may survive in MBL systems.
We numerically investigate a disordered Heisenberg chain which alters depending on the strength of the disordered magnetic filed \cite{pal10} between the thermalization and MBL phases. It shows that if a system thermalizes, it cannot be in a macroscopic superposition for long while a localized system can. 

Our study reveals an important role of the suppression of quantum fluctuation in the thermalization process and explains why a macroscopic quantum superposition is hardly observed in a real world even without the assumption of openness of quantum systems.
From this result, we also learn that a Hamiltonian which leads to thermalization is ``classical'' in the sense that it does not generate a superposition of macroscopically distinct states \cite{kofler08}. 

~

\textit{Measure of macroscopic superpositions.}-- 
There have been a number of studies on criteria and measures of macroscopic superpositions, and this type of nonclassical feature is often called ``quantum macroscopicity'' \cite{leggett80,shimizu02,shimizu05,lee11,frowis12,nimmrichter13,arndt14,frowis15,jeong15,kang16,vedral16}.
Several suggestions have been made to quantify quantum macroscopicity of arbitrary spin systems for pure states~\cite{shimizu02} and more generally including  mixed states~\cite{shimizu05,frowis12,kang16}.
Remarkably, for an arbitrary multipartite pure state $\ket{\psi}$, all those proposals \cite{shimizu02,shimizu05,frowis12,kang16} are reduced to the maximum variance of observables $A$ as
\begin{align}
	\mathcal{M}(\ket{\psi})	&= \max_{A \in S} \mathcal{V}_A(\ket{\psi}) \label{eq:def}
\end{align}
where $S$ is a set of sums of local spin operators with arbitrary directions and $\mathcal{V}_A(\ket{\psi}) = \braket{\psi|A^2|\psi}-\braket{\psi|A|\psi}^2$ is the variance of  $A$.
We consider a large system composed of  $N$  spin-1/2 particles, where $S$ is constructed as
\begin{align}
	S = \Bigl\{A=\sum_{i=1}^{N}  O_i
	 : |\vec{\alpha}_i| = 1 \text{ for all } 1\leq i \leq N \Bigr\} \label{eq:S}
\end{align}
where  $O_i = \vec{\alpha}_i \cdot \pmb{\sigma}^{(i)}$ is  a local spin operator for  the $i$-th particle and  $\pmb{\sigma}^{(i)}=\{\sigma^{(i)}_x,\sigma^{(i)}_y,\sigma^{(i)}_z\}$ is a Pauli spin operator applied to the $i$-th  particle. We shall call these observables $A$,
the sums of local observables, as macroscopic observables following Ref.~\cite{poulin05}. 
Although it is highly nontrivial to precisely measure the maximum variance of a multipartite spin system, for example, a recently proposed method to measure the variance of staggered magnetization in a trapped-ion system \cite{smith16} can be employed to find out the lower bound of quantum macroscopicity of MBL systems \cite{SM}.

Recently, the experimental measurability of quantum macroscopicity has been investigated for spin and bosonic systems \cite{jeong14,frowis2017}.
The interference-based measure for bosonic systems \cite{lee11}
can be  approximately measured in an optical setup using an overlap measurement~\cite{hendrych03} and local decoherence without a full tomography \cite{jeong14}.
Practically, this requires a beam splitter with two photon-number parity measurements to implement an overlap measurement and two additional beam splitters to implement local photon loss~\cite{jeong14}.
Cold atoms in an optical lattice may provide a more scalable setup to observe dynamics of macroscopic superpositions for bosonic systems.
Here, we note that the quantum macroscopicity measures for bosonic systems \cite{lee11,frowis15,oudot15,volkoff15} are also directly related to the maximum variances for pure states \cite{SM}. 
An atomic homodyne scheme may then be used to detect quadrature variables  \cite{gross11,tiesinga13} and estimate their maximum variances.
An alternative method using an overlap measurement~\cite{alves04,greiner15} with an appropriate unitary operator or a collective particle loss may also be considered for this purpose \cite{SM}.
Fr\"owis {\it et al.} investigated detectable lower bounds of the quantum Fisher information that are directly applicable in this context \cite{frowis2016,frowis2017}.

A quantum state $\ket{\psi}$ of $N$ spin-$1/2$ particles has a value of $\mathcal{M}(\ket{\psi})$ between $N$ to $N^2$~\cite{frowis12,kang16}.
A state $\ket{\psi}$ is not a macroscopic superposition if $\mathcal{M}(\ket{\psi}) = \mathcal{O}(N)$, using the big-O notation,
while $\mathcal{M}(\ket{\psi}) = N^2$ means that it has the maximum value of  quantum macroscopicity~\cite{shimizu02,frowis12,kang16}.
There are a number of justifying arguments behind this type of measure. They include the phase space structure~\cite{lee11,kang16}, sensitivity to  decoherence~\cite{shimizu02,kang16,lee11} and usefulness for quantum metrology~\cite{frowis12}. 
For example, measure $\mathcal{M}(\ket{\psi})$ can be understood as fragility of a quantum state. If $\mathcal{M}(\ket{\psi})>{\cal O}(N)$, the state becomes extremely fragile for a sufficiently large $N$ regardless of the coupling strength between the system and environment~\cite{shimizu02}. This is an anomalous situation for a classical system and implies the state is in a macroscopic superposition for $N\gg1$~\cite{shimizu02}.

Here, we add even another supporting argument based on coarse-grained measurements that $\mathcal{M}(\ket{\psi}) = \mathcal{O}(N)$ means absence of quantum macroscopicity.
A pure quantum state can be considered macroscopic 
when a coarse-grained measurement can give plural distinct outcomes   \cite{kofler08}.
Consider a 
coarse-grained measurement of $A$ with a finite resolution $\Delta$, within which a certain number of eigenvalues are contained, that cannot distinguish each eigenvalue $m$.
When the maximum variance of a macroscopic observable, $\max_{A\in S} \mathcal{V}_A(\ket{\psi})$, is larger than the resolution, the coarse-grained measurement that maximizes $\mathcal{V}_A(\ket{\psi})$ will give plural distinct results. 

It was argued that a product state of $N$ identical microscopic pure states $\ket{\phi}^{\otimes N}$
gives a single measurement outcome   under a coarse-grained measurement of  $\Delta^2>{\cal O}(N)$,
reproducing consistent results with classical physics~\cite{poulin05}.
It is straightforward to obtain $\mathcal{M}(\ket{\phi}^{\otimes N})=N$.
Given that a product state  $\ket{\phi}^{\otimes N}$ has no  quantum macroscopicity,
we can say that any state $\ket{\psi}$ with $\mathcal{M}(\ket{\psi}) = \mathcal{O}(N)$ has no quantum macroscopicity.
In fact, any product state gives the same result of $\mathcal{M}(\ket{\psi}) = \mathcal{O}(N)$ and
it is straightforward to show that a spin coherent state of $N$ spin-$1/2$ particles is another such example~\cite{kofler07}.

~

\textit{Systems under thermalization.}-- 
When a closed quantum system thermalizes, the eigenstate thermalization hypothesis (ETH) \cite{deutsch91,srednicki94,rigol08} can be employed in order to predict physical quantities after thermalization. The ETH states that a physical observable represented using the eigenbasis of the Hamiltonian has smooth diagonal components in energy. If the ETH is satisfied, we can relate the thermal ensemble averaged value and the time-averaged value of an observable.
We first assume that each local spin operator, $\sigma^{(i)}_a$, and each two-body operator, $\sigma^{(i)}_a \sigma^{(j)}_b$, where $i,j \in \{1,\cdots,N\}$ and $a,b \in \{x,y,z\}$ satisfy the ETH. Under this assumption, the time averaged values of $A$ and $A^2$ are obtained as~\cite{SM}
\begin{align}
\overline{A} &= \sum_i \braket{O_i}_T + \mathcal{O}(1) = \braket{A}_T + \mathcal{O}(1),\label{eq:Aave}\\
\overline{A^2} &= \sum_{i,j} \braket{O_i O_j}_T + \mathcal{O}(N) = \braket{A^2}_T+ \mathcal{O}(N), \label{eq:A2ave}
\end{align}
where $\overline{A}=\lim_{\tau\rightarrow \infty} \tau^{-1} \int_0^\tau A(t) dt$ indicates the long time averaged value of $A$ and $\braket{A}_T$ is the canonical thermal ensemble averaged value over temperature $T$ defined as $\braket{A}_T=\Tr[e^{-H/(k_B T)} A]/\Tr[e^{-H/(k_B T)}]$
with Hamiltonian $H$ and the Boltzmann constant  $k_B$. Here, the temperature $T$ for the canonical ensemble average can be obtained from equation  $\braket{\psi|H|\psi}=\braket{H}_T$ for initial state $\ket{\psi}$.

Using Eqs.~(\ref{eq:Aave}) and (\ref{eq:A2ave}), we obtain the following relation~\cite{SM}
\begin{align}
	\overline{\mathcal{V}_A}(\ket{\psi(t)}) &= 	\braket{A^2}_T - \braket{A}_T^2 + \mathcal{O}(N) \label{eq:qf},
\end{align}
where $\overline{\mathcal{V}_A}(\ket{\psi(t)}) $ is the time-averaged quantum fluctuation and
$\braket{A^2}_T - \braket{A}_T ^2$ corresponds to the thermal fluctuation.
When a system is in the thermal equilibrium, the thermal fluctuation should be suppressed over the size of the system, i.e., $(\braket{A^2}_T - \braket{A}_T ^2)/N^2 \rightarrow 0$ as $N$ increases. 
In fact, for a typical non-critical system, $\braket{A^2}_T - \braket{A}_T ^2$ behaves extensively so that it is $\mathcal{O}(N)$~\cite{landau80}. 
For instance, a one-dimensional non-critical short-range interacting system has a finite thermal correlation length $\xi$, and the order of the correlation function is $\braket{O_i O_j}_T - \braket{O_i}_T\braket{O_j}_T \sim \mathcal{O}(e^{-|i-j|/\xi})$. 
Equation~\eqref{eq:qf} then becomes $\overline{\mathcal{V}_A}(\ket{\psi(t)}) =\mathcal{O}(N)$ noting that $\braket{A^2}_T - \braket{A}_T^2 = \sum_{i,j}[\braket{O_i O_j}_T - \braket{O_i}_T\braket{O_j}_T]$ from Eq.~\eqref{eq:S}.
It immediately implies that ${\cal M}(|\psi(t)\rangle)= \mathcal{O}(N)$ after thermalization. In other words, relaxed states after thermalization are no longer macroscopic superpositions even though initial states were macroscopic superpositions.
The Hamiltonian of such a system generally does not produce a macroscopic superposition and can be regarded as a classical process~\cite{kofler08}.

~

\textit{MBL systems.}-- 
When a system is fully MBL, all eigenstates are localized and the system does not thermalize. 
A fully MBL system is completely characterized by local integrals of motion and the Hamiltonian can be written using those operators~\cite{serbyn13,huse14}.
The choice of local integrals of motion is not unique but we here follow the ``l-bits'' representation~\cite{huse14},  which is convenient for a MBL system of $N$ spin-$1/2$ particles, to investigate the lower bound of $\mathcal{M}(\ket{\psi(t)})$ for $t \rightarrow \infty$ \cite{SM}.
In this representation, local integrals of motion are generated from $N$ independent pseudospin operators $\tau^i_z$ which commute one another and with the Hamiltonian.
Each $\tau^i_z$  acts only on the localized region near site $i$, and it is thus related to the Pauli operators by a quasi-local unitary transform $U_{q}$ as $\tau^i_z = U_{q} \sigma^{(i)}_z U_{q}^\dagger$. The other pseudospin operators, $\tau^i_x$ and $\tau^i_y$, are defined accordingly.

The effective Hamiltonian can then be written in terms of localized Pauli operators $\{\tau^i_z\}$ as~\cite{serbyn13,huse14}
\begin{align}
	H_{\rm eff} = \sum_i \mathcal{E}_i \tau^i_z + \sum_{i,j} V_{i,j} \tau^i_z \tau^j_z + \cdots \label{eq:mbl_ham}
\end{align}
where we can neglect higher order terms in a deep localized regime. 
To calculate the lower bound of quantum macroscopicity, we consider an initial state $\ket{\psi_0}$ and a macroscopic observable $A = \sum_i O_i \in S$ where each $O_i$ is a local spin operator and has an overlap with $\tau^i_z$. In other words, we write $O_i = \gamma_i \tau^i_z + \cdots$ where the ellipsis contains first order of $\tau^i_x,\tau^i_y$ and higher order terms.
Then, we can obtain $\mathcal{V}_A(e^{-iHt}\ket{\psi(t)}) \approx \sum_{i, j} \gamma_i \gamma_j[\braket{\tau^i_z\tau^j_z} - \braket{\tau^i_z}\braket{\tau^j_z}]$ where $\ket{\psi(t)} = e^{-i H t} \ket{\psi_0}$ for $t \gg 1$.
We here neglected (i) the terms containing $\tau^i_x$ and $\tau^i_y$ as they decay following the power law~\cite{serbyn14} in MBL systems and (ii) the high order terms of $\tau^i_z$ as they are small in deep localized regime.
Using the original Pauli basis and appropriately choosing $A$, we obtain $\mathcal{M}(\ket{\psi(t)}) \geq c^2 \max_B [\mathcal{V}_B(\ket{\psi_0})]$ where $c  = \min_i |\vec{\beta}_i|^2$. 
The maximum is taken over macroscopic observables, $B = \sum_i (\pm \hat{\beta}_i )\cdot \pmb{\sigma}^{(i)}$, with all the possible combinations of the $\pm$ signs.
Here, the direction of the local spin operator $\hat{\beta}_i = \vec{\beta}_i/|\vec{\beta}_i|$ is determined by $\vec{\beta}_i = \{\Tr[\tau^i_z \sigma^{(i)}_x]/2,\Tr[\tau^i_z \sigma^{(i)}_y]/2,\Tr[\tau^i_z \sigma^{(i)}_z/2]\}$ and this can be calculated once $\tau^i_z$ is known.
Hence, we can see that $\mathcal{M}(\ket{\psi(t)})>O(N)$ even for $t \gg 1$ if $\max_B[\mathcal{V}_B(\ket{\psi_0})] > O(N)$.
A detailed analysis is presented in the appendix~\cite{SM}.

\textit{Numerical  analysis of disorderd Heisenberg chain.}-- 
We consider the Heisenberg spin-1/2 chain model with random fields along the $z$ direction. The Hamiltonian of the system is given by 
\begin{align}
	H_h=\sum_{i=1}^{N}[{\cal J} \pmb{s}^{i}\cdot \pmb{s}^{i+1} + h_i s^i_z + \Gamma s^i_x] 
\end{align}
where $\cal J$ is the interaction strength between two neighboring spins, $\pmb{s}^i=\pmb{\sigma}^{(i)}/2$ is a canonical spin-$1/2$ variable for the $i$-th spin, and $\{h_i\}$ and $\Gamma$ describe disorder and transverse fields, respectively.
Each $h_i$ is an independent random variable  picked up from a uniform distribution of $[-h,h]$ to describe the disorder field in the $z$ direction, 
and the transverse field $\Gamma$ in the $x$ direction breaks the total symmetry of $J_z=\sum_i \sigma^z_i$. 
We impose a periodic boundary condition $\pmb{s}^{N+1}=\pmb{s}^1$ and set ${\cal J}=1$ and $\Gamma=0.1$ for our numerical calculations.
Such a small field of $\Gamma\ll\cal J$ does not break the MBL phase transition~\cite{yang16}.
When $\Gamma \ll \mathcal{J}$, the system is known to thermalize for small positive $h$ and enters into the MBL phase when $h$ increases. 
Numerical studies with $\Gamma =0$ showed that some eigenstates start to localize for $h \gtrsim 2.0$ and the whole eigenstates are localized for $h > h_c \approx 3.6$ in the sector of $J_z=0$ for the finite $N$~\cite{pal10,luitz15,serbyn15}. MBL phases are robust against small local perturbations and the small field of $\Gamma=0.1 \cal J$ chosen in our study does not make significant changes~\cite{yang16}.

\begin{figure}[t]
	\centering
	\resizebox{0.46\textwidth}{!}
	{
		\includegraphics{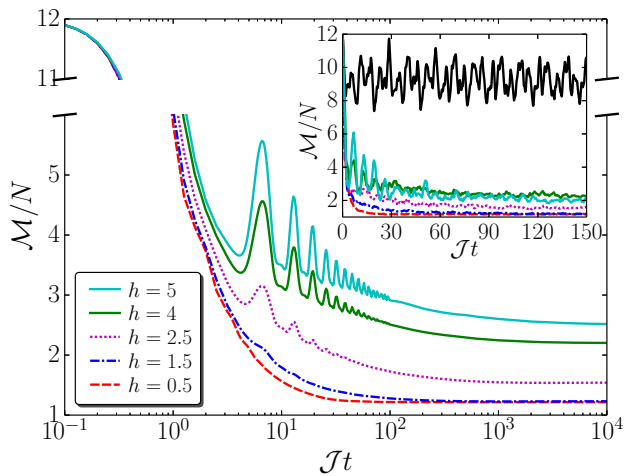}
	}
	\caption{ \label{fig:tvsi} (Color online)  Time evolution of normalized quantum macroscopicity.
	For the disordered Heisenberg spin chain ($\mathcal{J}_z = \mathcal{J}_\perp=1.0$) of size $N=12$, we take average of $\mathcal{M}(\ket{\psi(t)})/N$
	over many randonly chosen disorder realizations and initial GHZ states. We then plot the results of normalized quantum macroscopicity against time $t$.
	The dashed, dot-dashed, and dotted curves indicate the results for $h=0.5$, $h=1.5$, and $h=2.5$, respectively, for which the eigenstates are fully delocalized ($h=0.5$ and $h=1.5$) or partially localized ($h=2.5$). The results for $h=4$ (lower) and $h=5$ (upper) are indicated by the solid curves, which are the cases of the MBL phase. Obviously, $\mathcal{M}/N$ for the MBL phase converges to larger values than that for the thermalization phase does. 
		Inset: 
${\mathcal M}/N $ for a single initial state and single disorder realization. The uppermost black curve corresponds to the disordered XX model with $h=5$ (single-particle localized). 
}
\end{figure}

\begin{figure}[t]
	\centering
	\resizebox{0.46\textwidth}{!}
	{
		\includegraphics{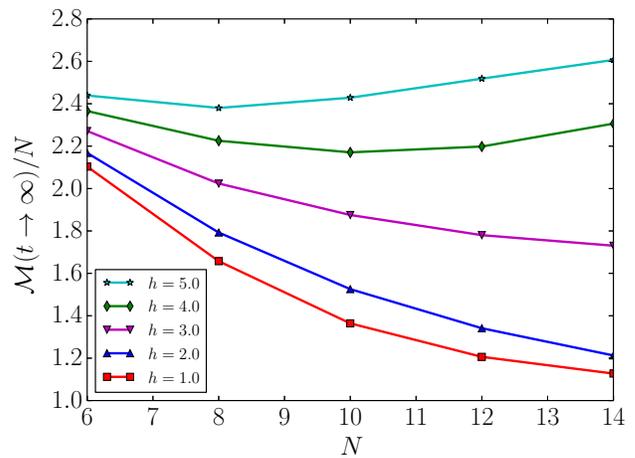}
	}
	\caption{\label{fig:nvsi} (Color online) Saturated values of normalized quantum macroscopicity for system sizes. The averaged saturated values of $\mathcal{M}/N$ against the number of particles $N$ are plotted. In thermal and intermediate regions ($h < h_c\approx 3.6$), quantum macroscopicity of initial GHZ states is not preserved and $\mathcal{M}/N$ approaches 1 as $N$ increases, which means $\mathcal{O}(N)$ behavior of $\mathcal{M}$. In contrast, we observe that $\mathcal{M}/N$ keeps increasing as $N$ increases (i.e., $\mathcal{M}>\mathcal{O}(N)$) for the MBL phase ($h > h_c$).}
\end{figure}

We fully diagonalize the Hamiltonian to characterize the time evolution of the system. We averaged over 10000 realizations of $\{h_i\}$ for $N=6$, 1000 realizations for $N=8$ and $N=10$, and 200 realizations for $N=12$ and $N=14$.
For each realization of $\{h_i\}$, we choose 100 random GHZ (Greenberger-Horne-Zeilinger) states for initial states and averaged all results. Each random GHZ state is constructed using $U_1\Motimes\cdots\Motimes U_N\ket{\mathrm {GHZ}_N}$ where each $U_i$ is a random unitary transform (in terms of Haar measure) in $\mathrm{SU}(2)$ and $\ket{\mathrm {GHZ}_N}=(\ket{\uparrow}^{\otimes N} + \ket{\downarrow}^{\otimes N})/\sqrt{2}$ is a GHZ state in $z$ direction \cite{SM}.

In Figs.~1 and 2, we display $\mathcal{M}/N$ instead of $\mathcal{M}$ to normalize the minimum value to $1$ and the maximum  to $N$, which is more intuitive and useful to investigate the scaling behaviors. In addition, we set $\hbar=1$ for simplicity. The average values of $\mathcal{M}/N$ for all realizations against  time $t$ for the system size $N=12$ are plotted in Fig.~\ref{fig:tvsi}. 
The starting value of $\mathcal{M}/N$ in Fig.~\ref{fig:tvsi} is $12$ because $\mathcal{M}=N^2$ for a GHZ state. 
To obtain the numerically optimized values of $\mathcal{M}$, the Broyden-Fletcher-Goldfarb-Shanno algorithm \cite{fletcher87} was used. 

Figure~\ref{fig:tvsi} shows that the saturated values of $\mathcal{M}/N$ for $t\gg1$ in the MBL phase ($h > h_c$) are larger than those in $h < h_c$.
The oscillations of $\mathcal{M}/N$ shown for $h \gtrsim 2.0$ in a certain time range are due to the oscillations of the spin correlation functions when the eigenstates are localized~\cite{serbyn14}. 
We further analyze the dynamics of $\mathcal{M}/N$ using the effective model of a MBL system ~\cite{serbyn13,huse14} and compared it with a single-particle localized system~\cite{SM}. 
For a single realization of disorder and an initial state, we plot the dynamics of $\mathcal{M}/N$ for the disordered Heisenberg model in Eq.~\eqref{eq:mbl_ham} with different values of $h$ and the disordered XX model with $h=5.0$, which is single-particle localized, in the inset of Fig.~\ref{fig:tvsi}. 
The results clearly show a sharp difference between MBL and single-particle localized systems; $\mathcal{M}/N$ converges as time evolves in a MBL system by dephasing but it tends to permanently oscillate and preserve large values in a single-particle localized system. 

The average saturated values of $\mathcal{M}/N$ as a function of $N$ are plotted in Fig.~\ref{fig:nvsi}, 
where the values decrease as $N$ increases for $h < h_c$. This means that $\mathcal{M}(\ket{\psi(t\gg1)}) = \mathcal{O}(N)$ for $h < h_c$ and thus the initial random GHZ states in the thermalization phase have lost their properties as macroscopic quantum superpositions for $t\gg1$.
On the other hands, $\mathcal{M}/N$ increases as $N$ increases in the cases of $h=4$ and $h=5$. This means that initial GHZ states in the MBL phase remain as macroscopic quantum superpositions for $t\gg1$.
However, as it was mentioned above, this behavior may depend on the initial condition of $\max_B [\mathcal{V}_B(\ket{\psi_0})]$. In the appendix~\cite{SM}, we show that there are some initial macroscopic superpositions that disappear in time, i.e., $\mathcal{M}(\ket{\psi(t\gg1)}) = \mathcal{O}(N)$, even though a system is in the MBL phase because the value of $\max_B[\mathcal{V}_B(\ket{\psi_0})]$ is small.
For a special type of initial states given by $\ket{\psi_0}=\exp(-i \sum_n \sigma^{(n)}_y \theta/2)^{\otimes N}\ket{\Psi_0}$, where $\ket{\Psi_0} = (\ket{\uparrow\downarrow\uparrow\downarrow\cdots}+\ket{\downarrow\uparrow\downarrow\uparrow\cdots})/\sqrt{2}$ and $\theta$ is an arbitrary angle, it would be feasible to experimentally detect large quantum macroscopicity in the MBL phase for $t \gg 1$ using trapped-ion systems~\cite{smith16,SM}.

We note that the decreasing behavior of $\mathcal{M}/N$ over $N$ in the thermalization phase does not conflict with the volume law of entanglement entropy which a thermal closed system should follow. Our measure $\mathcal{M}/N$ captures a different kind of quantumness that can be small even when entanglement is large~\cite{tichy16}.
In addition, since the onset of quantum chaos is directly related to validity of the ETH~\cite{santos10chaos}, it will be an interesting future work to investigate relations between arising of quantum chaos and disappearance of quantum macroscopicity in an isolated system.

\textit{Remarks.}--
It is well known that a macroscopic superposition rapidly disappears  due to interactions with its environment, while
it is often believed 
to survive if it is ideally isolated. We have investigated macroscopic superpositions in a closed system under the thermalization and MBL phases using a well-established measure of quantum macroscopicity~\cite{shimizu02,frowis12,kang16}.
Under the ETH, we have shown that the value of the measure is the order of $N$ for general non-critical short-range interacting spin systems where $N$ is the number of particles in the system. This means that the state after thermalization is not a macroscopic superposition any longer.  In contrast, macroscopic superpositions in the MBL phase may survive for $t\gg1$.

We have also performed numerical analyses with a disordered Heisenberg spin chain varied by the strength of the disorder between the thermalization phase and the MBL phase without thermalization.
Our numerical results confirm that macroscopic superpositions disappear when a closed system thermalizes, while they may be preserved  in the MBL phase in which the system does not thermalize.
Our work unveils a previously unknown aspect of fragility of macroscopic quantum superpositions even as a closed system, and provides a useful clue for engineering large-size quantum systems.

\textit{Acknowledgements.}--
We thank Hyukjoon Kwon, Dr. Minsu Kang, Dr. Malte C. Tichy, Prof. Martin B. Plenio for invaluable discussions. 
 The calculations in this work were performed using Alice cluster system in Quantum Information Group at Hanyang University. This work was supported by the National Research Foundation of Korea (NRF) grant funded by the Korea government (MSIP) (No. 2010-0018295) and by the KIST Institutional Program (Project No. 2E26680-16-P025).

~

\pagebreak

\hrulefill
\begin{center}
\textbf{\large APPENDIX}
\end{center}
\twocolumngrid

\setcounter{equation}{0}
\setcounter{figure}{0}
\makeatletter
\renewcommand{\theequation}{S\arabic{equation}}
\renewcommand{\thefigure}{S\arabic{figure}}
\renewcommand{\bibnumfmt}[1]{[S#1]}
\renewcommand{\citenumfont}[1]{S#1}
\newcommand{\rulesep}{\unskip\ \vrule\ }

\subsection*{Quantum fluctuation of a macroscopic observable for thermalizing systems}
The eigenstate thermalization hypothesis (ETH) \cite{deutsch91,srednicki94} provides a general method to predict physical quantities when a closed quantum system thermalizes. Many studies on various many-body systems have shown that the ETH is satisfied for local operators or two-point correlation functions~\cite{rigol08,rigol09,rigol10,santos10,ikeda11,steinigeweg13,steinigeweg14,ikeda13,beugeling14,sorg14,kim14,beugeling15,ikeda15}. 
Using these results, here we show that the quantum fluctuations of a \textit{macroscopic observable} after thermalization would be the same as the thermal fluctuations up to ${\cal O}(N)$. This leads to the proof of Eq.~(5) in the main text.

We summarize equilibration and thermalization of a closed system (for a comprehensive review, see Ref.~\cite{gogolin15} and references therein). Let $\ket{\alpha}$ denote an eigenstate of Hamiltonian $\hat{H}$ with eigenvalue $E_\alpha$. We consider $\alpha$ as an integer index and assume that the eigenstates $\ket{\alpha}$ are sorted in ascending order by the value of energy $E_\alpha$. 
For an initial state $\ket{\psi(0)}=\sum_\alpha C_\alpha \ket{\alpha}$, the state at time $t$ is given by
\begin{align*}
	\ket{\psi(t)} = \sum_\alpha C_\alpha e^{-i E_\alpha t} \ket{\alpha},
\end{align*}
and the time-dependent expectation value of a local operator $O$ is
\begin{align*}
	\braket{O(t)} &= \braket{\psi(t)|O|\psi(t)}\\
	= & \sum_\alpha |C_\alpha|^2 \braket{\alpha|O|\alpha} + \sum_{\alpha\neq\beta} C_\alpha^* C_\beta e^{i(E_\alpha-E_\beta)t}\braket{\alpha|O|\beta}. \numberthis \label{eq:time_exp}
\end{align*}
We first assume non-degenerate energies and energy gaps, and
the time averaged value of $\braket{O(t)}$ is expressed as
\begin{align*}
	\overline{O} = \lim_{\tau \to \infty} \frac{1}{\tau} \int_{0}^{\tau} dt \braket{O(t)}
= \sum_\alpha |C_\alpha|^2 \braket{\alpha|O|\alpha} \numberthis \label{eq:tavg}
\end{align*}
using Eq.~(\ref{eq:time_exp}).
This value is usually expressed as $\Tr[\rho_D O]$ where $\rho_D = \sum_\alpha |C_\alpha|^2 \ket{\alpha}\bra{\alpha}$
is a diagonal ensemble.
The equilibration requires $\braket{O(t)}$ would be the same with $\overline{O}$ for most of time $t$ which means that the second term of Eq.~\eqref{eq:time_exp} is small for general $t > 0$. There are two different arguments leading to this result. The first argument is based on the typicality, which shows that the second term is very small for typical initial states~\cite{reimann08,linden09,short11}. 
This argument shows that the equilibration is naturally induced from the distribution of coefficients $\{C_\alpha\}$ for typical initial states. In this case, the equilibration is explained without the ETH.
On the other hands, the ETH states that $\braket{\alpha|O|\beta}$ is exponentially small in the system size when $\alpha\neq\beta$ and this leads to the smallness of the second term. This is one of the basic assumption of ETH~\cite{srednicki96} and tested numerically in Refs.~\cite{rigol08,rigol09,santos10,beugeling15}. In any cases, we can see that the equilibration generally occurs.

In addition to the equilibration, the thermalization requires the expectation values of the diagonal ensemble to be the same with the microcanonical or canonical ensemble.
For this requirements, the ETH assumes that the diagonal element in energy basis $\braket{\alpha|O|\alpha}$ varies slowly as a function of $E_\alpha$ where the difference between neighboring diagonal components $\braket{\alpha+1|O|\alpha+1}-\braket{\alpha|O|\alpha}$ is exponentially small in the size of the system.  

We then define the total energy $\bar{E} = \sum_\alpha |C_\alpha|^2 E_\alpha$.
The microcanonical ensemble average can be expressed as
\begin{align*}
	\braket{O}_{\rm mc} &= \frac{1}{\mathcal{N}_{\bar{E},\Delta E}} \sum_{|E_\alpha-\bar{E}|\leq \Delta E} \braket{\alpha|O|\alpha}
\end{align*}
where $\mathcal{N}_{\bar{E},\Delta E}$ is number of states in the energy window centered at $\bar{E}$ and width $\Delta E$. The summation is applied only for the energy eigenstates in this window defined by $\bar{E}$ and $\Delta E$. It is known that the diagonal ensemble Eq.~\eqref{eq:tavg} approaches the microcanonical ensemble average for $(\Delta E)^2 |O''(E)/O(E)| \ll 1$  where $O(E)$ is the expectation value of observable $\braket{\alpha|O|\alpha}$ as the function of energy $E_\alpha$. Note that $O(E)$ is a smooth function in the thermodynamic limit from the ETH.

Actual numerical tests on diagonal ensembles and microcanonical ensembles using local operators and few-body operators show that they differ only in the order of $D^{-1/2}$ where $D$ is the dimension of the full Hilbert space~\cite{rigol09,santos10,ikeda13,beugeling14,sorg14}. Therefore we can say $\overline{O}  = \braket{O}_{\rm mc} + \mathcal{O}(D^{-1/2})$.

The difference between the diagonal ensemble and the canonical ensemble averaged values for theses observables also have been investigated. Under the ETH assumption, the difference between two ensembles for observables which do not depend on the size of the system (intensive observables) is given by $\mathcal{O}(1/N)$ where $N$ is the size of the system~\cite{srednicki99}. More direct comparison between two ensembles using numerical calculations for Bose-Hubbard model can be found in Ref.~\cite{sorg14}. 

We then calculate the long time averaged value of $A$ and $A^2$ where a macroscopic observable $A$ is given by
\begin{align*}
	A = \sum_i \vec{\alpha}_{i} \cdot \pmb{\sigma}^{(i)} = \sum_i O_i
\end{align*}
where $\vec{\alpha}_i$ are unit vectors and local operators $O_i=\vec{\alpha}_{i} \cdot \pmb{\sigma}^{(i)}$ are defined for simplicity. 
We now assume that all local spin operators $\sigma^{(i)}_a$ and two body observables $\sigma^{(i)}_a \sigma^{(j)}_b$ for $i,j \in \{1,\cdots,N\}$ and $a,b \in \{x,y,z\}$ satisfy the ETH.
Under these assumptions, we obtain Eqs.~(3) and (4), i.e., the time averaged values of $A$ and $A^2$ as
\begin{align*}
	\overline{A}=\overline{\braket{\psi(t)|A|\psi(t)}} & = \sum_{i} \overline{O_i}= \sum_{i}\bigl[ \braket{O_i}_{\rm mc}  + \mathcal{O}(D^{-1/2})\bigr]\\
	&= \sum_{i} \bigl[\braket{O_i}_T + \mathcal{O}(1/N)\bigr]\\
	&= \sum_{i} \braket{O_i}_T + \mathcal{O}(1) \\
	&=  \braket{A}_T + \mathcal{O}(1)
	\numberthis
\end{align*}
and
\begin{align*}
	\overline{A^2}&=\overline{\braket{\psi(t)|A^2|\psi(t)}}
	 = \sum_{i,j} \overline{O_i O_j} \\
	&= \sum_{i,j} \bigl[\braket{O_i O_j}_{\rm mc} + \mathcal{O}(D^{-1/2})\bigr]\\
	&= \sum_{i,j} \bigl[\braket{O_i O_j}_T + \mathcal{O}(1/N)\bigr]\\
	&= \braket{A^2}_T + \mathcal{O}(N). \numberthis 
\end{align*}

The  difference between $\overline{\braket{\psi(t)|A|\psi(t)}^2}$ and $\overline{A}^2$ is also needed to calculate the quantum fluctuation $\overline{\mathcal{V}_A}(\ket{\psi(t)})$.
We notice that this value is small because
\begin{align*}
	\overline{\braket{\psi(t)|A|\psi(t)}^2} - \overline{A}^2 &= \lim_{\tau \to \infty}\frac{1}{\tau} \int_0^\tau \bigl(\braket{A(t)} - \overline{A}\bigr)^2 \\
 &=  \sum_{\alpha\neq \beta}|C_\alpha|^2 |C_\beta|^2 |\braket{\alpha|A|\beta}|^2
\end{align*}
where the last expression is the averaged time fluctuation which is small when the equilibration occurs~\cite{linden09,reimann08,short11,srednicki96}.
Finally, we obtain the relation between the quantum fluctuation and the thermal fluctuation
\begin{align*}
	\overline{\mathcal{V}_A}(\ket{\psi(t) }) &= \overline{\braket{\psi(t)|A^2|\psi(t)}} - \overline{\braket{\psi(t)|A|\psi(t)}^2} \\
	&= \braket{A^2}_T - \braket{A}_T^2 + \mathcal{O}(N)
\end{align*}
which we used in Eq.~(5).

\subsection*{Quantum macroscopicity in localized systems}
When a system is fully localized, many local integrals of motion arise and they characterize the whole system. There are many different ways to choose the integrals of motions and their representations~\cite{serbyn13,huse14}. We here follow the ``l-bits'' representation \cite{huse14} that is convenient to investigate dynamical properties of the system (see e.g. Refs.~\cite{nanduri14,serbyn14,chandran15}). In this representation, the local integrals of motion are described using pseudospin operators that are connected to the Pauli operators in the physical basis with a quasi-local unitary transform. 
It means that there are operators $\{\tau^i_z\}$ which commute with each others ($[\tau^i_z,\tau^j_z] = 0$ if $i\neq j$) and with the Hamiltonian ($[\tau^i_z,H] = 0$ for all $i$). In addition, there is a quasi-local unitary transform $U$ such that $\tau^i_z = U \sigma^{(i)}_z U^\dagger$. Using this operator $U$, we can also define $\tau^i_x$ and $\tau^i_y$.
After constructing local integrals of motion, the Hamiltonian of many-body localized (MBL) systems can be written using these operators as
\begin{align}
	H = \sum_i \mathcal{E}_i \tau^i_z + \sum_{i,j} V_{i,j} \tau^i_z \tau^j_z + \sum_{i,j,k} V_{i,j,k} \tau^i_z\tau^j_z\tau^k_z + \cdots,
\label{eq:vvv}
\end{align}
where each $\mathcal{E}_i$ is a single spin excitation energy and $V_{i,j}$($V_{i,j,k}$) are two (three) body interaction strengths. The ellipsis contains high order terms which can be neglected in the deep localized regime.
The interaction terms in Eq.~(\ref{eq:vvv}) are a crucial property of a MBL system that is distinguished from the single particle localization (Anderson localization).
Many interesting features of the dynamics of MBL systems such as logarithmic increasing of bipartite entanglement~\cite{nanduri14}, dephasing effects in local density operators, and the power-law decay of temporal fluctuations of local spins~\cite{serbyn14} are originated from the interaction terms.
These characteristics of MBL systems are in contrast with single-particle localized systems that show properties such as freezing of bipartite entanglement and continuous fluctuating of local spin expectation values.

We then calculate a lower bound of $\mathcal{M}(\ket{\psi(t)})$ for $t \gg 1$ when a system is deeply many-body localized. Even though our derivation relies on several assumptions, it well describes behaviors of quantum macroscopicity of MBL systems as we shall see in the following sections.
From the completeness of $\{\tau^i_a\}$, a local operator $O_i = \vec{\alpha}_{i} \cdot \pmb{\sigma}^{(i)}$ can be expanded as
\begin{align}
O_i &= \sum_{\{k, a\}} \gamma^i_{\{k,a\}} \tau^{k_1}_{a_1}\tau^{k_2}_{a_2} \cdots \tau^{k_n}_{a_n} \nonumber \\
&= \sum_{k_1,a_1} \gamma^i_{k_1,a_1} \tau^{k_1}_{a_1} + \sum_{k_1,k_2,a_1,a_2} \gamma^i_{k_1,k_2,a_1,a_2} \tau^{k_1}_{a_1}\tau^{k_2}_{a_2}  + \cdots \label{eq:O_expand}
\end{align}
where the summation of $\{k, a\}$ runs over all $k=(k_1,k_2,\cdots,k_n) \subset \{1,2,\cdots,N\}$, $a=(a_1,a_2,\cdots,a_n)$. Here, $n=|k|$ and each $a_i \in \{x,y,z\}$ (see e.g. Ref.~\cite{serbyn14,chandran15} which used the same expansion).
The coefficients $\gamma^i_{\{k,a\}}$ are obtained using the orthogonality of $\{\tau^i_a\}$. For example, the coefficient for $\tau^i_z$ is given as $\gamma^i_{i,z} = \Tr[O_i \tau^i_z]/2$.
From the quasi locality of $U$, the coefficients $\gamma^i_{\{k,a\}}$ decay as
\begin{align}
\gamma^i_{\{k,a\}} \propto \exp[-\max(|k_\alpha - k_\beta|, |i - k_\alpha|)/\xi_2] \label{eq:gamma_order}
\end{align}
where the maximum is taken over all $\alpha,\beta \in \{1 \cdots n\}$ and $\xi_2$ is a characteristic length scale. 
To simplify the notation, we now take the Heisenberg picture and omit $\psi$ in bra-ket as $\braket{O(t)} = \braket{\psi(t)|O|\psi(t)}$. In addition, we define $T^{k}_{a} = \tau^{k_1}_{a_1}\tau^{k_2}_{a_2} \cdots \tau^{k_n}_{a_n}$. 
We then calculate the expectation value of $\braket{A(t)}$ for $A=\sum_i O_i$ as
\begin{align}
\braket{A(t)} &=  \sum_{i = 1}^N \braket{O_i (t)} \nonumber \\
&= \sum_{i = 1}^N \sum_{\{k,a\}} \gamma^i_{\{k,a\}} \braket{T^{k}_{a} (t)}.
\end{align}
Likewise, we obtain
\begin{align}
\braket{A^2(t)} &=  \sum_{i,j = 1}^N \braket{O_i (t) O_j(t)} \nonumber \\
&= \sum_{i,j = 1}^N \sum_{\{k,a\},\{l,b\}} \gamma^i_{\{k,a\}} \gamma^j_{\{l,b\}} \braket{T^{k}_{a} (t) T^{l}_{b}(t) }.
\end{align}
Using these expressions, the variance of $A$ is given as
\begin{align*}
\quad &\mathcal{V}_A(\ket{\psi(t)}) = \braket{A^2(t)} -\braket{A(t)}^2 \nonumber \\
&= \sum_{i,j = 1}^N \sum_{\{k,a\},\{l,b\}} \gamma^i_{\{k,a\}} \gamma^j_{\{l,b\}} \bigl[\braket{T^{k}_{a} (t) T^{l}_{b}(t)} \\
& \quad\quad\quad\quad\quad  -\braket{T^{k}_{a}(t)} \braket{T^{l}_{b}(t)} \bigr] \\
& = \sum_{i,j =1}^N \sum_ {a,b \in \{x,y,z\}} \gamma^i_{i,a} \gamma^j_{j,b} [\braket{\tau^i_a(t)\tau^j_b(t)} - \braket{\tau^i_a(t)}\braket{\tau^j_b(t)}] \\
& \quad + \sum_{i,j =1}^N \sum_{\substack{\{k,a\}, \{l,b\} \\ k \neq \{i\} \vee l \neq \{j\} } }  \gamma^i_{\{k,a\}} \gamma^j_{\{l,b\}} \times \\
& \quad\quad\quad\quad\quad \bigl[\braket{T^{k}_{a} (t) T^{l}_{b}(t)}  -\braket{T^{k}_{a}(t)} \braket{T^{l}_{b}(t)} \bigr]. \numberthis \label{eq:v_expansion} \\
\end{align*}
If a system is MBL, it is known that $\braket{T^{k}_{a}(t)} \propto 1/t^d$ if $\exists a_i \in \{x,y\}$ where $d$ is a power law exponent~\cite{serbyn14}. Thus, only the part for $a = z$ and $b = z$  contributes to the summation in the first term of Eq.~\eqref{eq:v_expansion}  when $t \gg 1$.
In addition, the second term of Eq.~\eqref{eq:v_expansion} consists of the operators with distance more than $1$ from $i$ and $j$. This term can be neglected when a system is deeply localized, $\xi_2 \ll 1$, as $\gamma^i_{\{k,a\}} \ll \gamma^i_{i,a}$ in this setting. 
To sum up, we obtain 
\begin{align}
\mathcal{V}_A(\ket{\psi(t)}) \approx \sum_{i,j =1}^N \gamma^i_{i,z} \gamma^j_{j,z} [\braket{\tau^i_z\tau^j_z} - \braket{\tau^i_z}\braket{\tau^j_z}]
\end{align}
in this limit.
We dropped $t$ from the right hand side as $\{\tau^i_z\}$ are constants of motion.

Using the basis of the original Pauli operators and the completeness of $\{\sigma^i_a \}$, we write
\begin{align*}
\tau^i_z &= \sum_{\{k,a\}} \kappa^i_{\{k,a\}} \sigma^{k_1}_{a_1}\sigma^{k_2}_{a_2}\cdots \sigma^{k_n}_{a_n} \\
&= \vec{\beta}_i \cdot \pmb{\sigma}^{(i)} + \sum_{\{k,a\}, k \neq \{i\}} \kappa^i_{\{k,a\}} \sigma^{k_1}_{a_1}\sigma^{k_2}_{a_2}\cdots \sigma^{k_n}_{a_n} \numberthis \label{eq:tau_expansion}
\end{align*}
where we took out  the term of $k = \{i\}$ from the summation and $\vec{\beta}_i = \{\Tr[\tau^i_z \sigma^{(i)}_x]/2,\Tr[\tau^i_z \sigma^{(i)}_y]/2,\Tr[\tau^i_z \sigma^{(i)}_z]/2\}$. 
The norm of this vector is $|\vec{\beta}_i| \leq \Tr[(\tau^i_z)^2]/2 = 1$.
In the deep localized regime, we can neglect the second term in Eq.~\eqref{eq:tau_expansion} and we obtain 
\begin{align*}
& \sum_{i,j =1}^N \gamma^i_{i,z} \gamma^j_{j,z} [\braket{\tau^i_z\tau^j_z} - \braket{\tau^i_z}\braket{\tau^j_z}] \\
& \approx \sum_{i,j =1}^N \gamma^i_{i,z} \gamma^j_{j,z} |\vec{\beta}_i||\vec{\beta}_j|[\braket{W_i W_j} - \braket{W_i}\braket{W_j}] \numberthis \label{eq:general_Wi}
\end{align*}
where $W_i = \hat{\beta}_i \cdot \pmb{\sigma}^{(i)}$ and $\hat{\beta}_i = \vec{\beta}_i/|\vec{\beta}_i|$ is a unit vector of $\vec{\beta}_i$. In addition, from the expansion of Eq.~\eqref{eq:O_expand} and the definition of $O_i$, we obtain $\gamma^i_{i,z} = \vec{\alpha}_i \cdot \vec{\beta}_i$.
To obtain the simplified expression of Eq.~\eqref{eq:general_Wi}, we find a constant $c\geq 0$ that a set of $N$ equations 
\begin{align}
(\vec{\alpha}_i \cdot \vec{\beta}_i)|\vec{\beta}_i| = \pm c, |\vec{\alpha}_i| = 1 \text{ for all } i \in 1,2,\cdots N \label{eq:c_equations}
\end{align}
has the solution of $\{\vec{\alpha}_i \}$. 
For this solution, the right hand side of Eq.~\eqref{eq:general_Wi} is given as $c^2 \mathcal{V}_B(\ket{\psi(t=0)})$ where $B = \sum_i \mathrm{sign}(\vec{\alpha}_i \cdot \vec{\beta}_i) W_i \in S$ is another macroscopic observable.
As we want to find a lower bound of the right hand side of Eq.~\eqref{eq:general_Wi}, it is worth considering the maximal possible value of $c^2$.
It is straightforward to show that the given set of the equations does not have a solution if $c > \min_i |\vec{\beta}_i|^2$  because $|\vec{\alpha}_i \cdot \vec{\beta}_i| \leq |\vec{\beta}_i|$. 
In addition, $c = \min_i |\vec{\beta}_i|^2$ yields a solution to Eq.~\eqref{eq:c_equations} as $\vec{\alpha}_i = \pm ( c/|\vec{\beta}_i|^2 ) \hat{\beta}_i + \vec{v}_i$ satisfies  the given set of the equations for any $\vec{v}_i \perp \vec{\beta}_i$ which makes $|\vec{\alpha}_i| = 1$.
Therefore, for this solution of $\{\vec{\alpha}_i \}$, we obtain the following lower bound:
\begin{align}
\mathcal{M}(\ket{\psi(t)}) \geq \mathcal{V}_A (\ket{\psi(t)}) \gtrsim c^2 \max_B \mathcal{V}_B(\ket{\psi(t=0)}) \label{eq:M_bound}
\end{align} 
where $A = \sum_i \vec{\alpha}_i \cdot \pmb{\sigma}^{(i)}$, $c = \min_i |\vec{\beta}_i|^2 \leq 1$ and the maximum is taken over for all possible choices of the signs for $B=\sum_i ( \pm \hat{\beta}_i) \cdot \pmb{\sigma}^{(i)}$. Here, the directions are given by $\vec{\beta}_i = \{\Tr[\tau^i_z \sigma^{(i)}_x]/2,\Tr[\tau^i_z \sigma^{(i)}_y]/2,\Tr[\tau^i_z \sigma^{(i)}_z]/2\}$.
We note that our lower bound depends on the value of $\mathcal{V}_B(\ket{\psi(t=0)})$ and it does not guarantee a large value of $\mathcal{M}(\ket{\psi(t)})$ for $t \gg 1$ even when $\mathcal{M}(\ket{\psi(t=0})$ is large if $\mathcal{V}_B(\ket{\psi(t=0)})$ is small. The examples for such cases are suggested in the second last section of this appendix.
In addition, we restricted our argument on the deep localized regime. Our bound is sufficient in the context of this study but we expect that it may be possible to obtain more general lower bounds.

\begin{figure}[t]
	\centering
	\resizebox{0.46\textwidth}{!}
	{
		\put(110,80){\makebox(0,0)[r]{\strut{}\Huge (a)}}
		\includegraphics{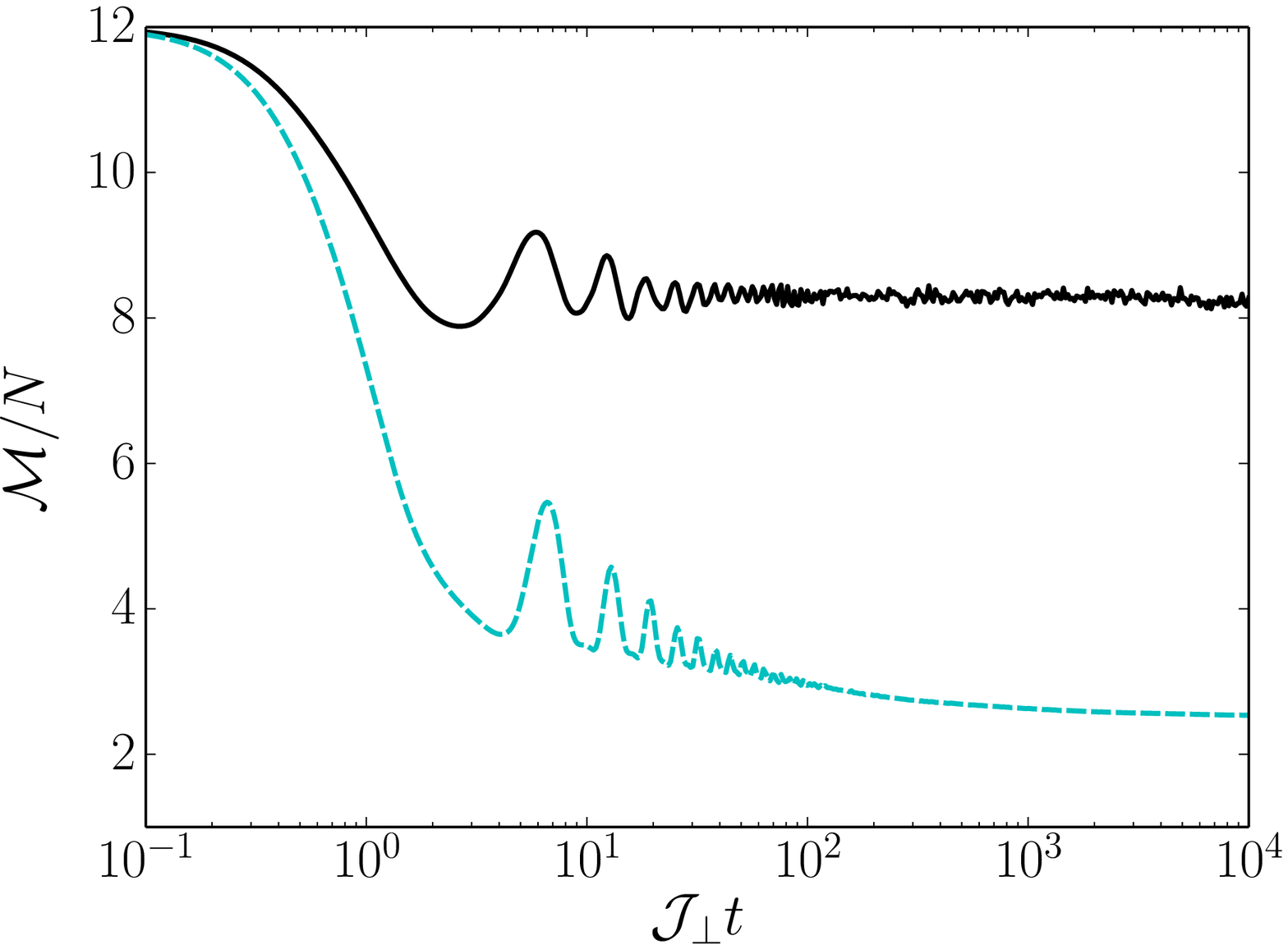}
	}\\
	\resizebox{0.46\textwidth}{!}
	{
		\put(110,80){\makebox(0,0)[r]{\strut{}\Huge (b)}}
		\includegraphics{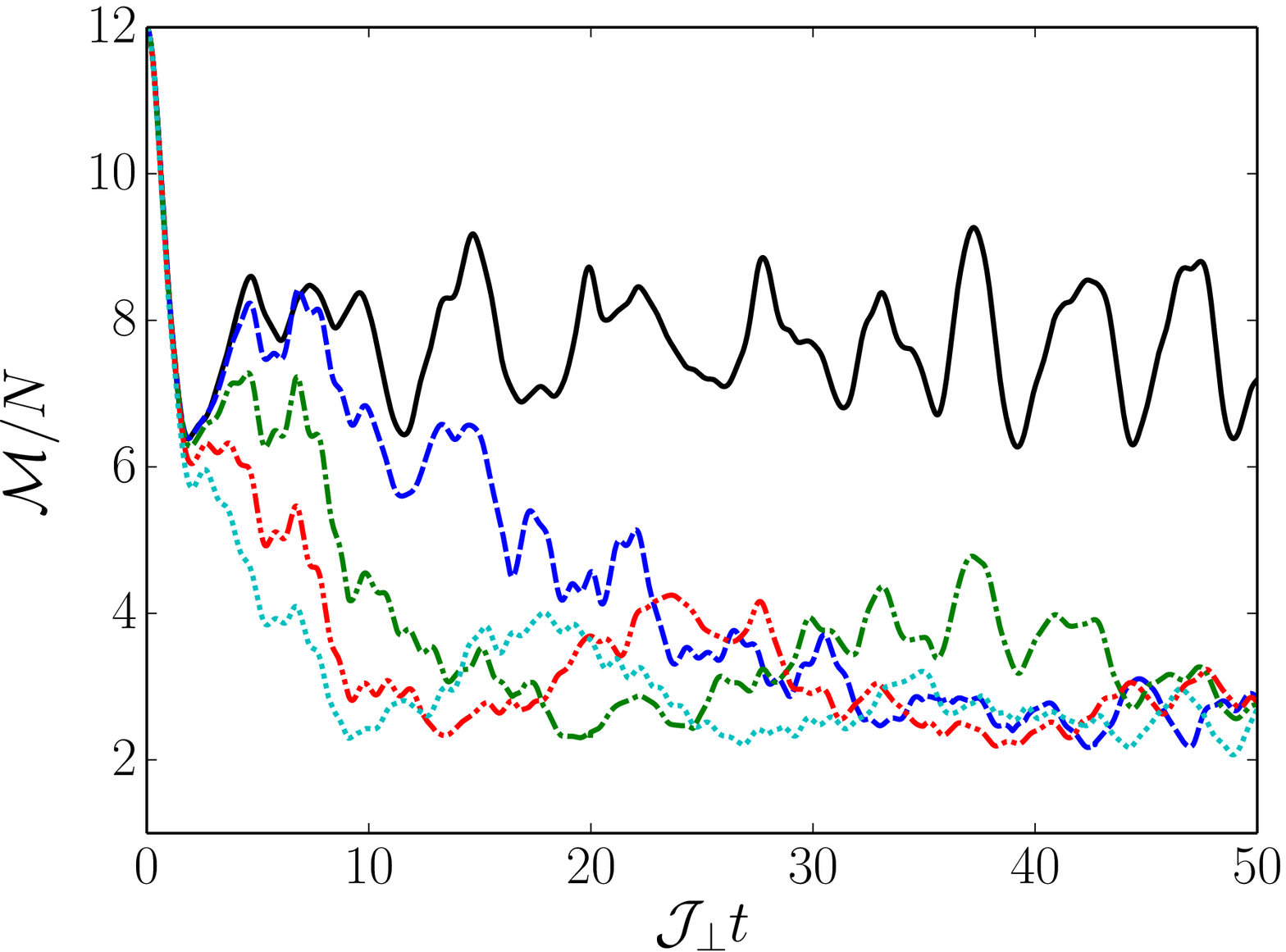}
	}
	\caption{\label{fig:mbl_vs_anderson} 
		(a) Averaged value of normalized quantum macroscopicity $\mathcal{M}(\ket{\psi(t)})/N$ over disorder realizations and initial states as a function of time. The solid curve shows the dynamics for the single particle localization case $\mathcal{J}_z = 0$ and $\Gamma = 0$ whereas the dashed curve corresponds to the dynamics for the MBL phase, $\mathcal{J}_z=1$ and $\Gamma=0.1$ (see the Hamiltonian in Eq.~\eqref{eq:ham}). Disorder strength $h=5$ is used for the both cases. 
		(b) Normalized quantum macroscopicity $\mathcal{M}(\ket{\psi(t)})/N$ for a single random GHZ initial state against time $t$. Different values of interaction strength $\mathcal{J}_z=0$, $0.1$, $0.2$, $0.3$, $0.4$ are indicated by solid, dashed, dot-dashed, dot-dot-dashed, and dotted curves, respectively.
		We set the disorder strength $h=5$ and the vanishing transverse field $\Gamma=0$. The dynamics of $\mathcal{M}/N$ for $\mathcal{J}_z = 0$ which corresponds to the single particle localization preserves the oscillation but it loses oscillating behavior for $\mathcal{J}_z > 0$.
	}
\end{figure}

In contrast, we consider a system that is single-particle localized with no interactions between pseudospins.
In this case, the correlation terms in Eq.~\eqref{eq:v_expansion} which contain $\tau^i_x$ or $\tau^i_y$ do not decay and oscillate permanently~\cite{serbyn14}. In other words, the terms for $a\neq z$ and $b\neq z$  in the first summation of Eq.~\eqref{eq:v_expansion} contribute to $\mathcal{V}_A(\ket{\psi(t)})$ even when $t \gg 1$. Therefore, we expect that quantum macroscopicity $\mathcal{M}(\ket{\psi})$ remains larger than the MBL case and the value continues to oscillate.

\begin{figure*}
	\resizebox{0.98\textwidth}{!}
	{
		\includegraphics{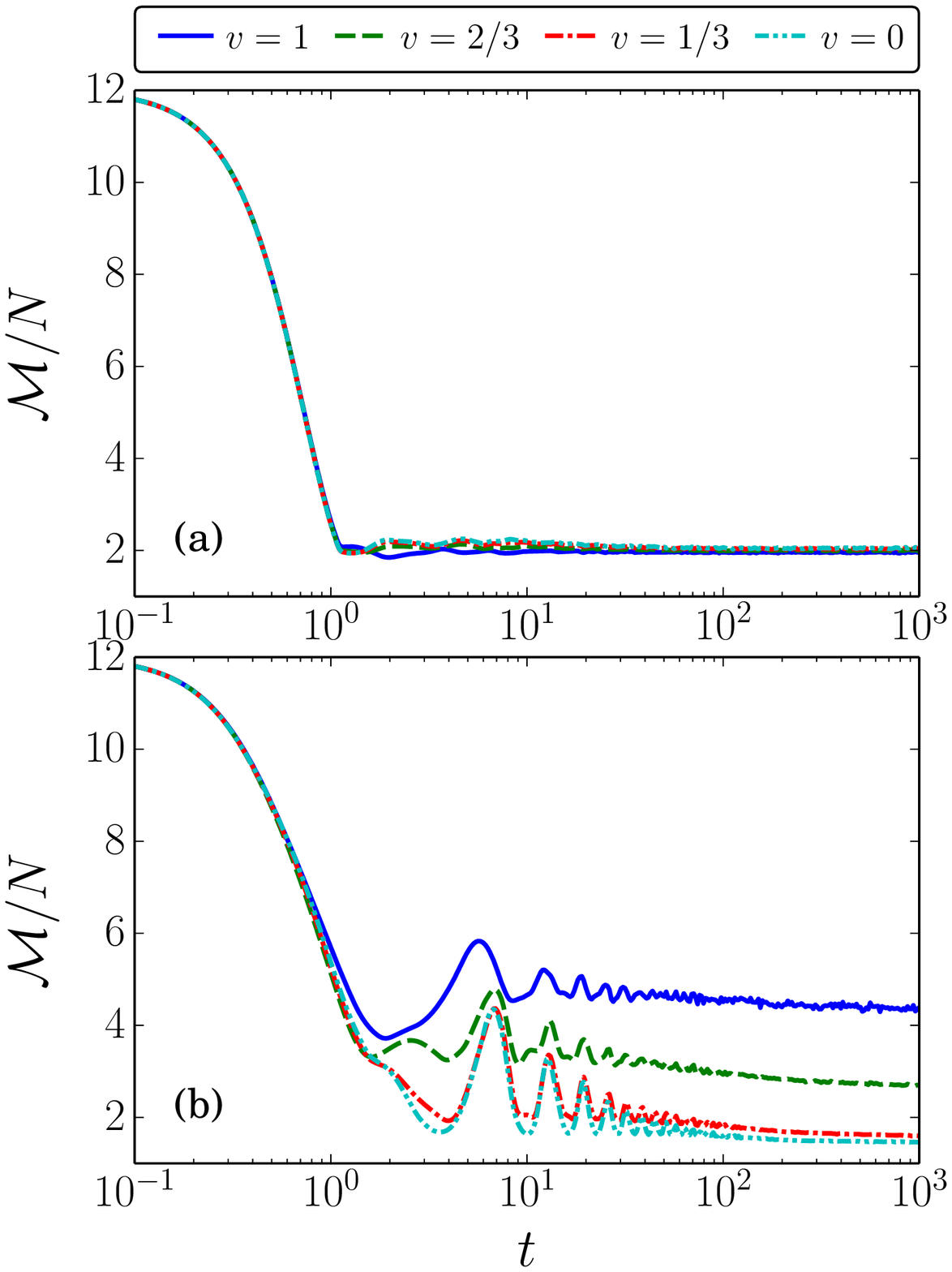}
		\hskip 1mm
		\vrule height 0.82\textheight width 2pt depth -0.08\textheight
		\includegraphics{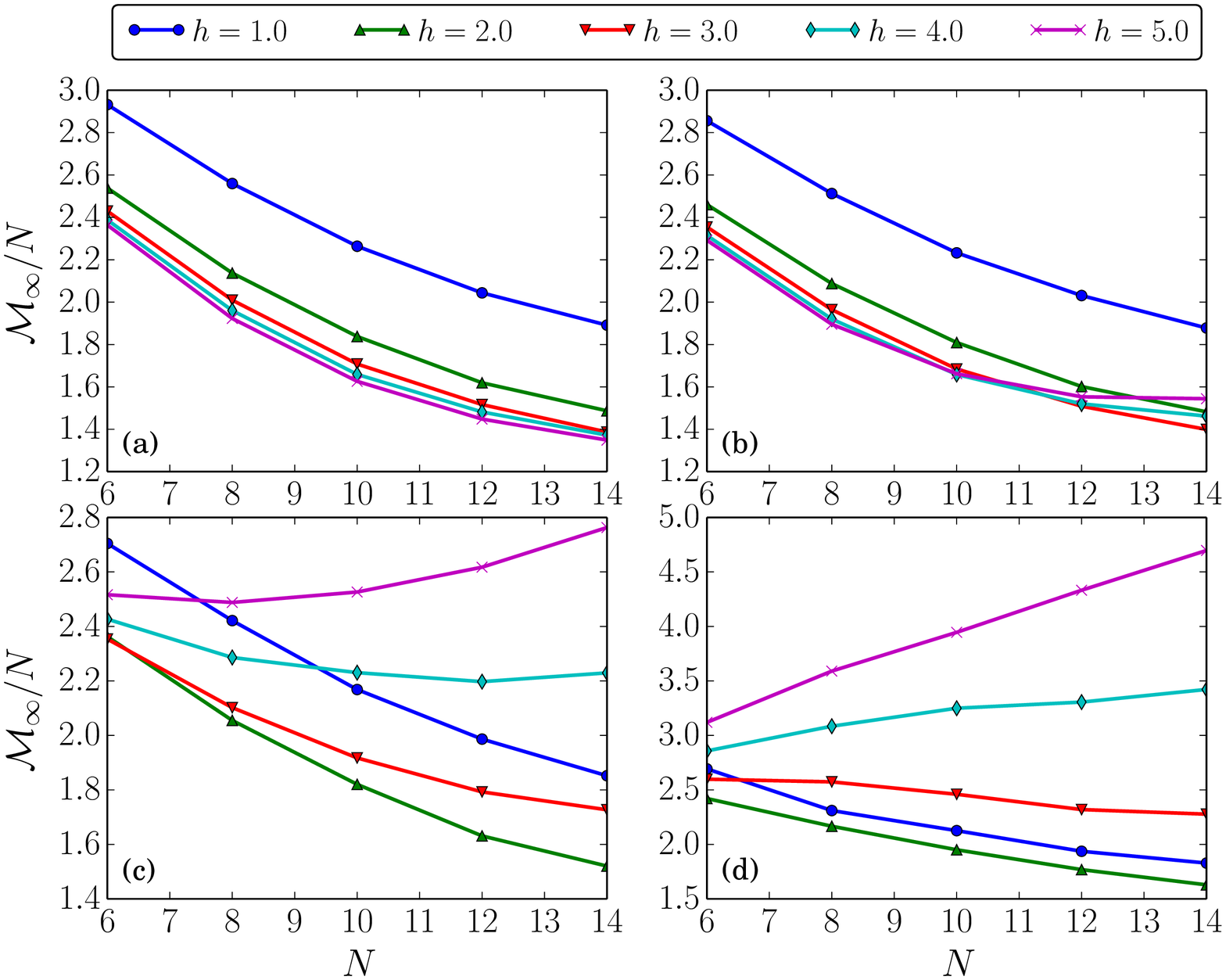}
	}
	\caption{\label{fig:vvsm}
		Left: Averaged values of normalized quantum macorscopicity $\mathcal{M}(\ket{\psi(t)})/N$ as functions of time $t$ for different values of $v=\cos\theta$ with the disorder strengths (a) $h = 1.0$ and (b) $h=5.0$. 
		Right: Averaged values of normalized quantum macorscopicity $\mathcal{M}(\ket{\psi(t)})/N$ at time $t \gg 1$ versus the size of system $N$ for (a) $v=0$, (b) $v=1/3$, (c) $v=2/3$, and (d) $v=1$. The results for different values from $h=1$ to $h=5$ are plotted.
	}
\end{figure*}

In order to verify this argument, we compare 
a single-particle localized system and a MBL system with the disorderd XXZ model the Hamiltonian
\begin{align}
	H = \sum_{i=1}^N \mathcal{J}_\perp (s^{i}_x s^{i+1}_x + s^{i}_y s^{i+1}_y) + \mathcal{J}_z s^{i}_z s^{i+1}_z + h_i s^i_z +\Gamma s^i_x \label{eq:ham}
\end{align}
where each $h_i$ is a random variable uniformly picked up from $[-h,h]$. The disordered Heisenberg model with a transverse field that we have considered in the main article is recovered when $\mathcal{J}_\perp = \mathcal{J}_z=\mathcal{J}$. When $\mathcal{J}_z = 0$ and $\Gamma = 0$ (for the disordered XX model), the Hamiltonian is directly mapped into non-interacting spinless fermions in a random potential $\sum_i c^\dagger_{i+1} c_i + h.c + h_i c_i^\dagger c_i$ using the Wigner-Jordan transformation which  represents a single-particle localized system. 
This Hamiltonian (\ref{eq:ham}) was also used for  the inset of Fig.~1 in the main article.

Figure~\ref{fig:mbl_vs_anderson}(a) shows the averaged values of $\mathcal{M}/N$ as time $t$ for these two different Hamiltonians for $N=12$. 
We average over the $100$ initial GHZ states in random local basis and $200$ disorder realizations.
The disorder strength $h=5.0$ and the hopping strength $\mathcal{J}_\perp = 1$ are used for both cases and we set $\hbar=1$ for convenience. In the case of the single-particle localization (solid curve), we see that the averaged value of $\mathcal{M}/N$ is preserved after initial oscillations whereas it decreases in the MBL phase (dashed curve) even after the oscillations that are a signature of slow dephasing by interactions.

The oscillations seem to disappear in time for both the cases of  the single-particle localization and the MBL phase
in Fig.~\ref{fig:mbl_vs_anderson}(a).
For the case of the single-particle localization (solid curve),
this disappearance is simply attributed to taking average over all realizations of disorders and initial states.
On the other hand, for the case  of the MBL phase (dashed curve), dephasing effects may cause the oscillations to disappear even for a single initial state.
In order to see this difference more clearly, we plot the normalized macroscopicity $\mathcal{M}/N$ for a single randomly generated initial GHZ state and a single disorder realization in Fig.~\ref{fig:mbl_vs_anderson}(b) with different values of $\mathcal{J}_z$. The disorder strength $h=5$ and the vanishing transverse field $\Gamma=0$ are used. It directly shows that the normalized quantum macroscopicity oscillates for a long time with a large amplitude and retains a larger value for $\mathcal{J}_z=0$.

\subsection*{Generating GHZ states in random local basis}
A random GHZ state can be constructed by applying local unitary transforms to the GHZ state in $z$ direction $\ket{\rm GHZ_N} = (\ket{\uparrow}^{\otimes N}+\ket{\downarrow}^{\otimes N})/\sqrt{2}$. 
To obtain random $\mathrm{SU}(2)$ operators, we use a parametrization of $U_i \in \mathrm{SU}(2)$ which is given by
\begin{align*}
	U_i = 
	\begin{pmatrix}
		e^{i\phi_i}\cos(\theta_i) & e^{i\xi_i}\sin(\theta_i) \\
		-e^{-i\xi_i}\sin(\theta_i)  &e^{-i\phi_i}\cos(\theta_i) \\
	\end{pmatrix}
\end{align*}
where $\phi_i,\xi_i \in [0,2\pi)$, $\theta_i \in [0,\pi]$. A Haar random unitary matrix $U_i$ is obtained by uniformly selecting $\phi_i,\xi_i \in [0,2\pi]$ and $\chi_i \in [0,1]$, and by setting $\theta_i = \arcsin \sqrt{\chi_i}$. Therefore, we can make an initial random GHZ state by computing $\ket{\psi(t=0)} = U_1\dots U_N \ket{\rm GHZ_N}.$

~

\subsection*{Initial state dependence}

We make further examinations of the dynamics and scaling behaviors of $\mathcal{M}/N$ using a different set of initial states. We here consider rotated N\'eel GHZ states as initial states which are given by $\ket{\psi(0)} = U^{\otimes N} \ket{\Psi_0}$ where $U=e^{-I \sigma_y \theta/2}$ is a rotation about the $y$ axis with angle $\theta$ and
\begin{align}
\ket{\Psi_0} = (\ket{\uparrow\downarrow\uparrow\downarrow\cdots}+\ket{\downarrow\uparrow\downarrow\uparrow\cdots})/\sqrt{2} \label{eq:neel_ghz}
\end{align}
 is a superposition between two different antiferromagnetic ordered N\'eel states.
We let $\cos\theta=v$ which corresponds to the $z$-coordinate of $U\ket{\uparrow}$ in the Bloch sphere. 
We consider four different values of $v$ which are given by $0, 1/3, 2/3$ and $1$ to set the initial states. 
The value $v=1$ means no rotation with $\theta = 0$, whereas $v=0$ corresponds to the states which are aligned over the $x$ axis with $\theta = \pi / 2$.

Using the Hamiltonian considered in the main article, which can be obtained from Eq.~\eqref{eq:ham} by setting $\mathcal{J}_\perp = \mathcal{J}_z = 1.0$ and $\Gamma = 0.1$, we calculate $\mathcal{M}/N$ as a function of time $t$ in Fig.~\ref{fig:vvsm} (Left). The size of system $N=12$ and four different values of $v$ are used. 
The results show that $\mathcal{M}/N$ approaches a particular value as time evolves in the thermalization phase ($h = 1.0$) regardless of the value of $v$. 
However, the converged values of $\mathcal{M}/N$ in the MBL phase ($h=5.0$) increase with $v$.

\begin{figure}[t]
	\resizebox{0.48\textwidth}{!}
	{
		\includegraphics{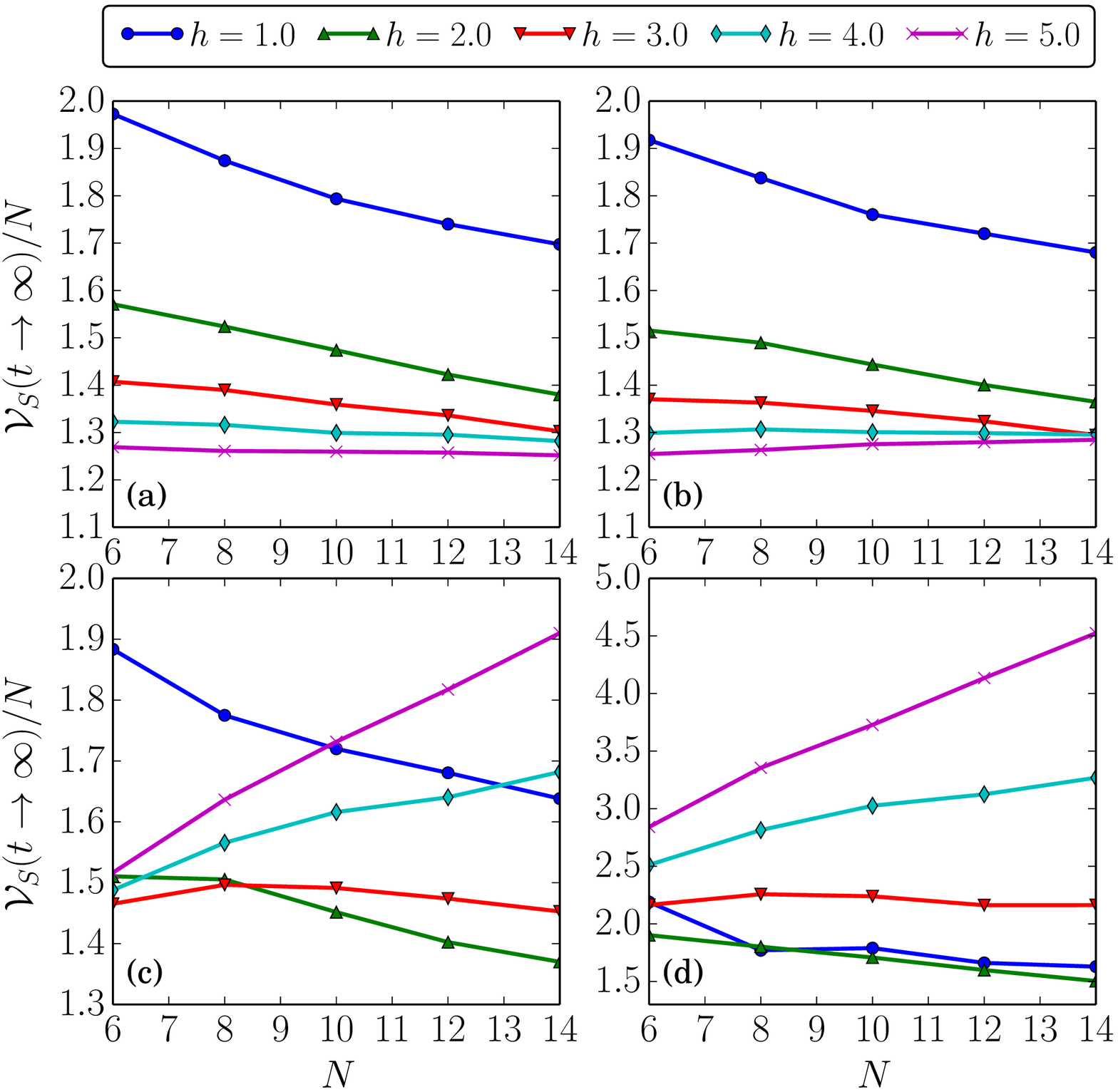}
	}
	\caption{\label{fig:vvsm_nvsm_smag}
		Averaged saturated values of the variance of staggered magnetization $S(\theta)$ in the rotated basis versus the size of system $N$. Four different values of (a) $v =0$, (b) $1/3$, (c) $2/3$, and (d) $1$ are used. Staggered magnetizations in the rotated basis $S(\theta)$ are sufficient to capture the behaviors of quantum macroscopicity $\mathcal{M}=\max_A \mathcal{V}_A$ in MBL systems.
	}
\end{figure}

We also calculated the values of $\mathcal{M}/N$ after the equilibration as functions of $N$ in Fig.~\ref{fig:vvsm} (Right). We can find that $\mathcal{M}/N$ does not increase with $N$ even for the MBL phase if the initial states are prepared with $v=0$ (for the case of (a)). 
The lower bound of Eq.~\eqref{eq:M_bound} may give a hint to this behavior. In the case of vanishing the transverse field, i.e., $\Gamma = 0$, the conservation of total spin-z operator $J_z = \sum_i \sigma^{(i)}_z$ makes $\hat{\beta}_i = \hat{z} = \{0,0,1\}$~\cite{serbyn14}.
We can then simply calculate $\max_B \mathcal{V}_B(\ket{\psi(t=0)}) = N + (N^2-N) \cos^2 \theta$
for combinations of $B = \sum_i (\pm) \sigma^{(i)}_z$. As we expect that the small value of $\Gamma=0.1$ does not make a significant difference, our lower bound indicates that the saturated values of $\mathcal{M}/N$ must increase as $N$ increases for $v > 0$ but it can retain a small value for $v=0$.

~

\subsection*{Experimental considerations}

There have been several experimental realizations of MBL systems~\cite{schreiber15,choi16,smith16}. We here focus on a trapped-ion implementation~\cite{smith16} as trapped ions provide high controllability in preparation of initial states and measurement of various operators.

First, we note that the measurement of staggered magnetization in the rotated basis gives the maximum variance for rotated N\'{e}el GHZ states. Explicitly, the operator
\begin{align}
S(\theta) = \sum_{i=1}^N (-1)^i \vec{\chi}_i \cdot \pmb{\sigma}^{(i)}
\end{align} 
where $\vec{\chi}_i = \{\sin\theta, 0, \cos\theta\}$ gives the maximum variance $\mathcal{V}_{S(\theta)}(U^{\otimes N} \ket{\Psi_0}) = N^2$. In addition, as $S(\theta)$ is a macroscopic observable, $\mathcal{M}(\ket{\psi}) \geq \mathcal{V}_{S(\theta)}(\ket{\psi})$.

We calculated the saturated values of the variance of the operator $S(\theta)$ for the initial rotated N\'{e}el states as the size of the system $N$ in Fig.~\ref{fig:vvsm_nvsm_smag} using the Hamiltonian in the main article. The results show the similar behavior with Fig.~\ref{fig:vvsm} (Right). Importantly, it shows the increasing behavior for $v=1/3,2/3,1$ as $N$ increases in the MBL phase. It means that we can see preservation of quantum macroscopicity in the MBL phase only using the operator $S(\theta)$.

We finally comment on how to prepare an initial state $U^{\otimes N} \ket{\Psi_0}$ and to measure $\mathcal{V}_{S(\theta)}$ in a trapped-ion system. 
First, the initial state of the form $U^{\otimes N} \ket{\Psi_0}$ can be prepared using trapped ions as GHZ states up to $N=14$ can be arranged~\cite{monz11} and single qubit rotations with extreme high fidelity ($\approx 99.9999\%$)~\cite{harty14} can be applied consecutively.
Subsequently, the system is subject to the disordered Ising Hamiltonian with long-range interactions~\cite{smith16} for $t \gg 1$. After that, we rotate all qubits into the original $z$ basis and measure the staggered magnetization in $z$ basis as in~\cite{smith16}. After repeating many measurements, the variance $\mathcal{V}_{S(\theta)}$ can be calculated.

However, it is not simple to answer whether macroscopic superpositions would disappear in a trapped-ion quantum simulator as long-range interactions are present in this system.
It requires a more detailed study to show that how long-range interactions affect the behavior of $\mathcal{V}_{S(\theta)}$.

~

\subsection*{Experimental measurability of quantum macroscopicity for bosonic systems}

Here, we here discuss measures of quantum macroscopicity for bosonic systems and how they can be measured for optical fields and for bosonic particles in optical lattices. 
A general measure for bosonic systems was  suggested in Ref.~\cite{lee11} by quantifying interference fringes in the phase space. For a single-mode pure state, the measure can be expressed as
\begin{align}
\mathcal{I}_{ LJ} =  \braket{\psi| \Delta x^2| \psi} + \braket{\psi| \Delta p^2| \psi} -1 
\end{align}
where $x = (a + a^\dagger)/\sqrt{2}$ and $p = -i(a - a^\dagger)/\sqrt{2}$ are quadrature operators and $a_i$ and $a_i^\dagger$ are bosonic annihilation and creation operators.
This quantity can be approximately measured using an overlap measurement  between two copies of a system and local decoherence~\cite{jeong14}.
In an optical setup, this requires a beam splitter to mix the two copies, two beam splitters to implement local photon loss for each of the copies, and two photon number parity measurements; this method is valid for arbitrary states including mixed states \cite{jeong14}.

Another type of measure to quantify quantum macroscopicity of bosonic systems
\cite{frowis15,oudot15,volkoff15} is based on the quantum Fisher information. It should be noted that this approach was first introduced for spin systems in Ref.~\cite{frowis12}.
For a multipartite pure state, it corresponds to  
\begin{align}
\mathcal{I}_{b} =  \max_{\theta_{1,...,N} \in \mathbb{R} } \braket{\psi| \Delta \bigl( \sum_{i=1}^N x_i^{\theta_i} \bigr)^2| \psi}
\label{eq:Ib}
\end{align}
where $x_i^{\theta_i} = (a_i e^{-i \theta_i }+a_i^\dagger e^{i \theta_i })/\sqrt{2}$ is the generalized quadrature operator for
particle $i$ with the annihilation (creation) operator $a_i$ ($a_i^\dagger $) and $N$ is the number of modes.
It is straightforward to notice that this measure is in the form of Eq.~(1) of the main article while the set of macroscopic observables $S$ is now
\begin{align}
S = \Bigl\{\sum_{i=1}^{N}  x_i^{\theta_i} : \theta_i \in [0, 2 \pi) \text{ for all } 1\leq i \leq N \Bigr\}.
\end{align}
We shall assume an optical lattice where bosonic atoms are distributed to $N$ sites in the lattice.
In order to measure $\mathcal{I}_{b}$ for this system, a direct measurement of quadrature $x_i^{\theta_i}$ at each site of the optical lattice can be considered. Like the case for measuring individual spins of trapped ions, one may obtain the variance of $\sum_i x_i^{\theta_i}$ by performing atomic homodyne measurements for each site \cite{gross11,tiesinga13}.


Alternatively, overlap measurements may be considered  for this purpose.
An overlap measurement between two quantum states can be performed in an optical lattice
because it is known that beam-splitter-like operations and parity measurements are possible, just as they can be done in an optical setup, in an optical lattice~\cite{alves04,greiner15}.
We recall that the overlap between states $\rho$ and $\sigma$ is given as $\Tr[\rho \sigma]$.
Suppose two copies of a system in state $\ket{\psi}$. After we apply a unitary operator $U = e^{-i x A}$ where $A = \sum_{i=1}^{N}  x_i^{\theta_i} \in S$ to one of the copies, we get $\rho = \ket{\psi}\bra{\psi}$ and $\sigma = e^{-i x A}\ket{\psi}\bra{\psi}e^{i x A}$. The overlap between two states is then $\Tr[\rho \sigma] = F(\ket{\psi},e^{-i \theta A} \ket{\psi})^2$ where $F(\ket{\psi},e^{-i x A} \ket{\psi}) = |\braket{\psi | e^{-i x A} | \psi}|$ is the fidelity between the two pure states. For a small value of $x$, the fidelity can be expressed as
\begin{align}
F(\ket{\psi},e^{-i x A} \ket{\psi}) = 1 - \frac{x^2}{2} \mathcal{V}_A (\ket{\psi}) + \mathcal{O}(x^3)
\end{align}
where $\mathcal{V}_A (\ket{\psi}) = \braket{\psi|A^2|\psi} - \braket{\psi|A|\psi}^2$ is the variance of $A$ for  state $\ket{\psi}$. One can then obtain the value of $\mathcal{V}_A (\ket{\psi})$, which is a lower bound for $\mathcal{I}_b$ using the unitary operator $e^{-i x A}$ and the overlap measurement. 
In an optical setup, the displacement operation can be implemented using a strong coherent state and a beam splitter \cite{banaszek99prl}. 
As $e^{-i x A}$ is a kind of the displacement operation and a superfluid ground state in an optical lattice is well approximated to coherent states~\cite{greiner2002}, one may attempt to implement the unitary operation $e^{-i x A}$ in optical lattices.

A different setup without the unitary operator may also be considered for overlap measurements. As in Ref.~\cite{jeong14}, a Lindblad decoherence channel can be employed to estimate values of $\mathcal{I}_b$. 
The overlap measurement on two copies of a quantum state $\rho$ will yield  $\Tr [\rho^2]$, i.e., the purity  of the state.
The purity decay rate $-\frac{d}{d t} \Tr [\rho^2]$ can also be obtained by applying a loss channel to both of the copies.
Using the Lindblad master equation with a single loss channel $L$ 
\begin{align}
\frac{ \partial \rho}{\partial t} = \left[L \rho L^{\dagger }- \frac{1}{2} \left(\rho L^{\dagger}L+L^{\dagger }L \rho \right)\right],
\end{align}
we obtain the purity decay rate as
\begin{align}
-\frac{d}{d t} \Tr [\rho^2] = 2 \Tr\left[\rho^2 L^{\dagger}L - \rho L \rho L^{\dagger }  \right].
\end{align}
The purity decay rate  for state $\ket{\psi}$ is then $-\frac{d}{d t} \Tr [\rho^2]=2 [\braket{\psi|L^\dagger L | \psi} - \braket{\psi| L^\dagger | \psi}\braket{\psi| L | \psi}]$. 
If there is a collective atomic loss with suitable phases given as $L = \sum_{i=1}^N a_i e^{-i \theta_i}$, we obtain
\begin{align*}
&-\frac{1}{2}\frac{d}{d t} \Tr [\rho^2] =\braket{\psi|L^\dagger L | \psi} -\braket{\psi| L^\dagger | \psi}\braket{\psi| L | \psi} \\
&\quad \quad = \sum_{i,j =1}^N [\braket{\psi|a_i^\dagger a_j | \psi} - \braket{\psi| a_i^\dagger | \psi}\braket{\psi| a_j | \psi} ] e^{i(\theta_i-\theta_j)} \\
&\quad \quad = \frac{1}{2} \bigl[\mathcal{V}_{\sum_i x_i^{\theta_i}}(\ket{\psi})+\mathcal{V}_{\sum_i x_i^{\theta_i + \pi/2}}(\ket{\psi}) - N\bigr] \numberthis \label{eq:decay_var}
\end{align*} 
for any values of $\theta_i$. This leads to the following inequalities:
\begin{align}
\mathcal{I}_b(\ket{\psi}) \leq \max_{\{ \theta_i\} } \bigl\{-\frac{d}{d t}\Tr [\rho^2] \bigr\} + N  \leq 2\mathcal{I}_b(\ket{\psi}),
\end{align} 
where the first inequality is obtained by assigning the optimal values of $\{\theta_i\}$ that maximize $\mathcal{I}_b$ in Eq.~(\ref{eq:Ib})  to Eq.~\eqref{eq:decay_var}. We get the second one straightforwardly by maximizing both sides of Eq.~\eqref{eq:decay_var}.
The inequalities imply that the purity decay rate gives both lower and upper bounds of $\mathcal{I}_b(\ket{\psi})$. In order to approximately obtain the purity decay rate, one needs to measure the purity $\Tr [\rho^2]$ before and after the loss \cite{jeong14}.

It may not be easy to experimentally realize the collective atomic loss channel $L=\sum_{i=1}^N a_i e^{-i \theta_i}$ in optical lattices. However, there have been many recent studies on experimental accessibility of various types of dissipations (see e.g. Ref.~\cite{muller12}). We expect that there could be a realizable scheme to implement the loss of $L=\sum_{i=1}^N a_i e^{-i \theta_i}$ in an optical lattice, although its further investigations are beyond the scope of this work.

~

~

\end{document}